\documentclass[12pt]{article}
\usepackage{amsmath,amsfonts,amssymb}
\usepackage[pdftex,colorlinks=false]{hyperref}

\makeatletter
\def\numberbysection{\@addtoreset{equation}{section}
    \def\theequation{\thesection.\arabic{equation}}}
\makeatother

\numberbysection

\newcommand{\be}{\begin{eqnarray}}
\newcommand{\ee}{\end{eqnarray}}
\newcommand{\non}{\nonumber}
\newcommand{\id}{\mathbb{I}}
\newcommand{\tr}{\mathop{\rm tr}\nolimits}
\newcommand{\diag}{\mathop{\rm diag}\nolimits}

\newcommand{\g}{\mathop{\mathfrak{g}}\nolimits}
\newcommand{\h}{\mathop{\mathfrak{h}}\nolimits}
\newcommand{\hh}{\mathop{\mathfrak{h'}}\nolimits}

\begin{document}

\begin{titlepage}
\strut\hfill UMTG--265
\vspace{.5in}
\begin{center}

\LARGE Nested algebraic Bethe ansatz\\
\LARGE for open $GL(N)$ spin chains\\
\LARGE with projected $K$-matrices\\
\vspace{1in}
\large Rafael I. Nepomechie \footnote{nepomechie@physics.miami.edu}\\[0.8in]
\large Physics Department, P.O. Box 248046, University of Miami\\[0.2in]  
\large Coral Gables, FL 33124 USA\\

\end{center}

\vspace{.5in}

\begin{abstract}
We consider an open spin chain model with $GL(N)$ bulk symmetry that
is broken to $GL(M) \times GL(N-M)$ by the boundary, which is a
generalization of a model arising in string/gauge theory.  We prove
the integrability of this model by constructing the corresponding
commuting transfer matrix.  This construction uses operator-valued
``projected'' $K$-matrices.  We solve this model for general values of
$N$ and $M$ using the nested algebraic Bethe ansatz approach, despite
the fact that the $K$-matrices are not diagonal.  The key to obtaining
this solution is an identity based on a certain factorization property
of the reduced $K$-matrices into products of $R$-matrices.  Numerical
evidence suggests that the solution is complete.
\end{abstract}

\end{titlepage}

\setcounter{footnote}{0}

\section{Introduction}\label{sec:intro}

As shown long ago by Sklyanin \cite{Sk}, the construction of an
integrable open spin chain model requires two main ingredients: an
$R$-matrix (solution of the Yang-Baxter equation) which determines the
bulk terms in the Hamiltonian, and right/left $K$-matrices (solutions
of the boundary Yang-Baxter equation \cite{Ch, GZ}) which determine
the right/left boundary terms in the Hamiltonian.  By now it is well
understood how to solve models with diagonal $K$-matrices.  However,
solving models with $K$-matrices which are {\it not} diagonal in
general remains a challenging problem.  Such $K$-matrices can have
matrix elements which are either $c$-numbers or operators.  While
models with non-diagonal $c$-number-valued $K$-matrices have received
considerable attention (see, for example, \cite{CLSW}-\cite{FNR}),
much less is known about models with operator-valued $K$-matrices.

An interesting class of non-diagonal operator-valued $K$-matrices
consists of so-called projected $K$-matrices found by Frahm and
Slavnov \cite{FS}.  Integrable open spin chains constructed with
$K$-matrices of this type have found applications in condensed matter
physics \cite{ZG, ZGLG, ZGLG2} and string/gauge theory \cite{BV, HM,
Ne2}.  \footnote{For related work in string/gauge theory, see e.g.
\cite{Mann}-\cite{Ga} and references therein.}

We consider here an integrable open spin chain model constructed with
such projected $K$-matrices.  The chain has $L+2$ sites, labeled $X$, 1,
\ldots, $L$, $Y$.  The space of states is
\be
\stackrel{\stackrel{X}{\downarrow}}{C^{M}}  \otimes 
\stackrel{\stackrel{1}{\downarrow}}{C^{N}}  \otimes \cdots
\stackrel{\stackrel{L}{\downarrow}}{C^{N}}  \otimes 
\stackrel{\stackrel{Y}{\downarrow}}{C^{M}} \,,
\label{Hilbertspace}
\ee
where $1<M<N$.  That is, the vector spaces of the ``bulk'' sites
(labeled 1, \ldots, $L$) all have dimension $N$, while the vector
spaces of left and right ``boundary sites'' (labeled $X$ and $Y$,
respectively) have a lower dimension $M$.  The Hamiltonian is given
by
\be
H = Q_{X}^{(M)} h_{X, 1} Q_{X}^{(M)} +\sum_{l=1}^{L-1} h_{l, l+1} + 
Q_{Y}^{(M)} h_{L, Y} Q_{Y}^{(M)}  \,,
\label{Hamiltonian}
\ee
where the two-site Hamiltonian $h_{l, l+1}$ is given by 
\be
h_{l, l+1} = \id_{l, l+1} - {\cal P}_{l, l+1} \,,
\label{twosite}
\ee
where $\id$ and ${\cal P}$ are the identity and permutation matrices 
on $C^{N} \otimes C^{N}$, respectively; and $Q^{(M)}$ is a diagonal
$N \times N$ matrix which projects 
$C^{N}$ to $C^{M}$,
\be
Q^{(M)} = \diag(\underbrace{1, 1, \ldots, 1}_{M}, \underbrace{0, 0, 
\ldots, 0}_{N-M}) \,.
\ee
We drop the null rows and columns of the left and right boundary terms
in the Hamiltonian, which therefore should be understood as $M N 
\times M N$ matrices acting on $C^{M} \otimes C^{N}$ and $C^{N}
\otimes C^{M}$, respectively.

Although the bulk terms have $GL(N)$ symmetry, the boundary terms
reduce the symmetry to $GL(M) \times GL(N-M)$.  We shall refer to this
model as the $GL(N)/(GL(M) \times GL(N-M))$ model.  The case
$(N,M) = (3,2)$ was recently studied (following \cite{BV, HM}) in
\cite{Ne2}.  \footnote{The idea of breaking a symmetry down to a
subgroup by boundary interactions has recently been explored a great deal in
the $O(N)$ case at the critical point in \cite{DJS}-\cite{BHK}.}

Within the quantum inverse scattering method, the standard approach
for solving integrable spin chains with higher-rank symmetry is nested
algebraic Bethe ansatz (ABA) \cite{KR, BdVV}.  This approach has been
adapted to open spin chains with diagonal $K$-matrices in \cite{FK,
dVGR, more}.  We further adapt this method, along the lines in
\cite{ZG, ZGLG, ZGLG2} for a related model with $M=2$, to solve the
$GL(N)/(GL(M) \times GL(N-M))$ model for general values of $N$ and
$M$.  The identity (\ref{remarkableM}), which relies on a certain
factorization property of the ``reduced'' $K$-matrices into products of
$R$-matrices, plays an essential role in obtaining a solution.
Numerical evidence suggests that the solution is complete.

The outline of this paper is as follows.  In Sec.  \ref{sec:transfer}
we construct the transfer matrix corresponding to the Hamiltonian
(\ref{Hamiltonian}), thereby proving the integrability of the latter.
In Sec.  \ref{sec:nested} we consider, as a warm-up, the special case
$M=2$.  We establish our notation, present the nested ABA solution,
and provide some evidence of its completeness.  We then treat the
general case $M\ge 2$ in Sec \ref{sec:nestedM}.  Finally, in Sec.
\ref{sec:conclusion} we present our conclusions and list some
interesting unresolved questions.  Appendix \ref{sec:proof} contains
our proof of the important identity (\ref{remarkableM}).

\section{Transfer matrix}\label{sec:transfer}

As already mentioned in the Introduction, the transfer matrix is
constructed from an $R$-matrix and right/left $K$-matrices.  The
former is a solution $R(u)$ of the Yang-Baxter equation (YBE)
\be
R_{12}(u_{1}-u_{2})\, R_{13}(u_{1})\, R_{23}(u_{2}) =
R_{23}(u_{2})\, R_{13}(u_{1})\, R_{12}(u_{1}-u_{2}) \,.
\label{YBE}
\ee 
In view of the $GL(N)$ symmetry of the bulk terms of the Hamiltonian, 
we take the well-known rational solution
\be
R(u) = u \id + i {\cal P} = a(u)  \sum_{a=1}^{N} e_{a a}^{(N)} \otimes 
e_{a a}^{(N)} + b(u) \sum_{a, b = 1\atop a\ne b}^{N} e_{a a}^{(N)} \otimes 
e_{b b}^{(N)} + i \sum_{a, b = 1\atop a\ne b}^{N} e_{a b}^{(N)} \otimes 
e_{b a}^{(N)}
\,,
\label{Rmatrix}
\ee
where 
\be
a(u) = u+i \,, \qquad b(u) = u \,,
\label{ab}
\ee 
and $e_{a b}^{(N)}$ is the standard elementary $N \times N$ matrix whose 
$(a, b)$ matrix element is 1, and all others are zero; i.e., 
$\big[e_{a b}^{(N)}\big]_{ij} = \delta_{a i} \delta_{b j}$.

The right $K$-matrix $K^{-}(u)$, which here acts on $C^{N} \otimes 
C^{M}$, is a solution of the right boundary Yang-Baxter equation 
(BYBE) \cite{Sk, Ch, GZ}
\be
\lefteqn{R_{12}(u_{1}- u_{2})\, K^{-}_{13}(u_{1})\, R_{12}(u_{1}+u_{2})\, 
K^{-}_{23}(u_{2}) }\non \\
& & = K^{-}_{23}(u_{2})\, R_{12}(u_{1}+u_{2})\, K^{-}_{13}(u_{1})\,R_{12}(u_{1}- 
u_{2}) \,.
\label{RBYBE}
\ee 
We take the solution \cite{FS} \footnote{See Eq. (3.12) for $\pi_{1} K_{-}(u) 
\pi_{1}$ in \cite{FS}. Here we take the 
constant $c=0$ (in order to match with the Hamiltonian
(\ref{Hamiltonian})), we rescale the rapidity $u \mapsto -i u$ (in
order to match with our conventions for the $R$-matrix
(\ref{Rmatrix})), and we clear the denominators by performing an overall rescaling.}
\be
K^{-}(u) = a_{1}(u)\sum_{a,b=1}^{M} e_{a a}^{(N)} \otimes e_{b 
b}^{(M)} + a_{2}(u)\sum_{a,b=1}^{M} e_{a b}^{(N)} \otimes e_{b 
a}^{(M)} + a_{3}(u)\sum_{a=M+1}^{N}\sum_{b=1}^{M} e_{a a}^{(N)} \otimes e_{b 
b}^{(M)} \,,
\label{KRsltn1}
\ee
where 
\be
a_{1}(u) = 1 - u^{2} \,, \qquad a_{2}(u) = - 2i u \,, \qquad a_{3}(u) = 1 + u^{2} 
\,.
\label{KRsltn2}
\ee
For the case $(N, M) = (3, 2)$, this matrix coincides with the one we
used earlier in \cite{Ne1}.  (There we called the left and right   
$K$-matrices $K^{L}$ and $K^{R}$ instead of $K^{+}$ and 
$K^{-}$; and we labeled the left and right spaces 0 and $L+1$ instead 
of $X$ and $Y$, respectively.)
Unfortunately, we were unaware of \cite{FS} at that time.

Here we define the left $K$-matrix $K^{+}(u)$ to also act on $C^{N} \otimes 
C^{M}$. \footnote{In \cite{Ne1}, we instead defined the left 
$K$-matrix to act 
on $C^{M} \otimes C^{N}$ (with $(N, M) = (3, 2)$); i.e., the two 
$K$-matrices are related by permutation.} It satisfies the left BYBE 
\cite{Sk}
\be
\lefteqn{R_{12}(-u_{1}+ u_{2})\, K^{+}_{13}(u_{1})^{t_{1}}\, 
R_{12}(-u_{1}-u_{2}-\eta)\, 
K^{+}_{23}(u_{2})^{t_{2}} }\non \\
& & = K^{+}_{23}(u_{2})^{t_{2}}\, R_{12}(-u_{1}-u_{2}-\eta)\, 
K^{+}_{13}(u_{1})^{t_{1}}\,R_{12}(-u_{1} +u_{2}) \,,
\label{LBYBE}
\ee 
where $t_{i}$ denotes transposition in the $i^{th}$ space, and 
$\eta = i N$ appears in the crossing-unitarity relation
\be
R_{12}(u)^{t_{1}} \, R_{12}(-u-\eta)^{t_{1}} \propto \id \,,
\ee 
where the proportionality factor is some scalar function of $u$.
A solution is provided by the ``less obvious'' isomorphism \cite{Sk}
\be
K^{+}_{13}(u) = \tr_{2} {\cal P}_{12} R_{12}(-2u-\eta) K^{-}_{23}(u) 
\,,
\ee
which gives (up to an irrelevant overall factor)
\be
K^{+}(u) = b_{1}(u)\sum_{a,b=1}^{M} e_{a a}^{(N)} \otimes e_{b 
b}^{(M)} + b_{2}(u)\sum_{a,b=1}^{M} e_{a b}^{(N)} \otimes e_{b 
a}^{(M)} + b_{3}(u)\sum_{a=M+1}^{N}\sum_{b=1}^{M} e_{a a}^{(N)} \otimes e_{b 
b}^{(M)} \,,
\label{KLsltn1}
\ee
where 
\be
b_{1}(u) = u^{2} +i(N-M) u\,, \qquad b_{2}(u) = 2i u - N\,, \qquad 
b_{3}(u) = -u^{2} - i M u \,.
\label{KLsltn2}
\ee
Again, for the case $(N, M) = (3, 2)$, this solution agrees with the one 
used in \cite{Ne1}. 

The transfer matrix $t(u)$ is given by \cite{Sk}
\be
t(u) = \tr_{a} K^{+}_{a X}(u)\, T_{a 1 \cdots L}(u) \, K^{-}_{a Y}(u)\,
\hat T_{a 1 \cdots L}(u)   \,,
\label{transfer}
\ee
where the trace $( \tr )$ is over an $N$-dimensional auxiliary space denoted by 
$a$. The argument of the trace acts on 
\be
\stackrel{\stackrel{a}{\downarrow}}{C^{N}}  \otimes 
\stackrel{\stackrel{X}{\downarrow}}{C^{M}}  \otimes
\stackrel{\stackrel{1}{\downarrow}}{C^{N}}  \otimes \cdots
\stackrel{\stackrel{L}{\downarrow}}{C^{N}}  \otimes 
\stackrel{\stackrel{Y}{\downarrow}}{C^{M}} \,,
\ee
and therefore $t(u)$ acts on (\ref{Hilbertspace}), as does the
Hamiltonian.  The monodromy matrices $T$ and $\hat T$ are given by
\be
T_{a 1 \cdots L}(u) = R_{a 1}(u) \cdots R_{a L}(u) \,, \qquad
\hat T_{a 1 \cdots L}(u) = R_{a L}(u) \cdots R_{a 1}(u) \,.
\label{monodromy}
\ee 
Indeed, it can be shown that the transfer matrix (\ref{transfer})
obeys the fundamental commutativity property
\be
\left[ t(u) \,, t(v) \right] = 0 \,.
\label{commutativity}
\ee
It can also be shown that this transfer matrix contains the
Hamiltonian (\ref{Hamiltonian}),
\be
H = c_{1} \frac{d}{du}t(u) \Big\vert_{u=0} + c_{2} \id \,,
\label{tHrelation}
\ee
where
\be
c_{1} =  \frac{i}{2N} (-1)^{L} \,, \qquad 
c_{2} = L + 1 + \frac{1}{N} \,.
\label{tHcoeffs}
\ee
The relations (\ref{commutativity}) - (\ref{tHcoeffs}) demonstrate 
the integrability of the Hamiltonian.

The transfer matrix has the $GL(M) \times GL(N-M)$ 
symmetry
\be
\left[ t(u) \,, \h \otimes \g^{\otimes L} \otimes \h \right] = 0 \,,
\ee
where 
\be
\g = \left( \begin{array}{cc}
                   \h & 0 \\
		    0 & \hh \\
	 \end{array} \right)  \,, \qquad 
	 \h \in GL(M)\,,  \quad \hh \in GL(N-M)  \,.
\label{g}
\ee

\section{Nested ABA for $M=2$}\label{sec:nested}

We now proceed to diagonalize the transfer matrix of the $GL(N)/(GL(M)
\times GL(N-M))$ model via nested ABA for the special case $M=2$.

\subsection{Preliminaries}\label{subsec:prelim}

We begin by assembling the ingredients needed to carry out the ABA
analysis: suitable operators, pseudovacuum states and commutation
relations.
For the $M=2$ case, the left $K$-matrix (\ref{KLsltn1}) has the form
(as an $N \times N$ matrix in the auxiliary space)
\be
K^{+}_{a X}(u) = \left(
 \begin{array}{cccccc}
     \alpha_{11}(u) &\alpha_{12}(u) & 0 & 0 &\cdots & 0\\
     \alpha_{21}(u) &\alpha_{22}(u) & 0 & 0 &\cdots & 0\\
                        0 & 0 & \beta(u)  & 0 &\cdots & 0\\
			0 & 0 & 0  & \beta(u) &\cdots & 0\\
     \vdots    & \vdots  & \vdots  &\vdots &\ddots  & \vdots  \\
     	0 & 0 & 0  & 0 &\cdots & \beta(u)
 \end{array} \right)  \,,
\label{KLform}
\ee
where $\alpha_{jk}(u)$ and $\beta(u)$ are operators on the 
two-dimensional quantum space $X$.
For future reference, we now introduce a ``down'' pseudovacuum state
for this space,
\be
|0\rangle_{X} =  
\left( \begin{array}{c}
0 \\
1 \end{array} \right) \,,
\label{0pseudovac}
\ee
and note that it is an eigenstate of the diagonal operators,
\be
\alpha_{11}(u) |0\rangle_{X} &=&  b_{1}(u) |0\rangle_{X} \,, \non \\
\alpha_{22}(u) |0\rangle_{X} &=&  (b_{1}(u) +b_{2}(u)) |0\rangle_{X} \,, \non \\
\beta(u) |0\rangle_{X} &=&  b_{3}(u) |0\rangle_{X} \,,
\label{diag0vac}
\ee 
and is annihilated by $\alpha_{12}(u)$,
\be
\alpha_{12}(u) |0\rangle_{X} = 0 \,.
\label{a120vac}
\ee

The right $K$-matrix (\ref{KRsltn1}) has a similar structure.
Introducing a ``down'' pseudovacuum state also for the quantum space
$Y$,
\be
|0\rangle_{Y} =  
\left( \begin{array}{c}
0 \\
1 \end{array} \right) \,,
\label{Lplus1pseudovac}
\ee
we see that it is an eigenstate of the diagonal operators 
\footnote{We denote by $\left[K^{-}(u)\right]_{jk}$ the $(j,k)$
element of $K^{-}_{a Y}(u)$ considered as an $N \times N$ matrix in the
auxiliary space, analogously to (\ref{KLform}).}
\be
\left[K^{-}(u)\right]_{11} |0\rangle_{Y} &=& a_{1}(u) |0\rangle_{Y}\,, \non \\
\left[K^{-}(u)\right]_{22} |0\rangle_{Y} &=& (a_{1}(u)+a_{2}(u)) |0\rangle_{Y}\,, \non \\
\left[K^{-}(u)\right]_{jj} |0\rangle_{Y} &=& a_{3}(u) |0\rangle_{Y}\,, \quad j = 3,\ldots, N \,,  
\label{kRdiagvac}
\ee
and is annihilated by $\left[K^{-}(u)\right]_{12}$,
\be
\left[K^{-}(u)\right]_{12} |0\rangle_{Y} = 0 \,.
\label{kR12vac}
\ee

The transfer matrix (\ref{transfer}) can be reexpressed as
\be
t(u) = \tr_{a} K^{+}_{a X}(u)\, {\cal T}^{-}_{a 1 \cdots L Y}(u) \,,
\label{transfer2}
\ee
where ${\cal T}^{-}_{a 1 \cdots L Y}(u)$, defined by 
\be 
{\cal T}^{-}_{a 1 \cdots L Y}(u) = T_{a 1 \cdots L}(u) \, K^{-}_{a Y}(u)\,
\hat T_{a 1 \cdots L}(u) \,,
\label{calT}
\ee
also obeys the right BYBE (\ref{RBYBE}).  It is from this object that
we must identify suitable operators (among them, creation-like
operators).  In view of the form (\ref{KLform}) of the left
$K$-matrix, we follow \cite{ZG, ZGLG, ZGLG2} (see also \cite{GM, FK,
dVGR, more} and references therein) and write ${\cal T}^{-}_{a 1
\cdots L Y}(u)$ as follows (as an $N \times N$ matrix in the auxiliary
space)
\be
{\cal T}^{-}_{a 1 \cdots L Y}(u)  = \left(
 \begin{array}{cccc}
     A_{11}^{(1)}(u) &\cdots & A_{1, N-1}^{(1)}(u) & B_{1}^{(1)}(u)\\
     \vdots    &\ddots &\vdots       & \vdots  \\
     A_{N-1, 1}^{(1)}(u) &\cdots & A_{N-1, N-1}^{(1)}(u) & B_{N-1}^{(1)}(u)\\
     C_{1}^{(1)}(u)  &\cdots & C_{N-1}^{(1)}(u)  &D^{(1)}(u)
 \end{array} \right)  \,,
 \label{calTform}
\ee
where $A_{jk}^{(1)}(u)\,, B_{j}^{(1)}(u)\,,  C_{j}^{(1)}(u)\,, D^{(1)}(u)$ are operators on
the quantum spaces
\be
\stackrel{\stackrel{1}{\downarrow}}{C^{N}}  \otimes \cdots
\stackrel{\stackrel{L}{\downarrow}}{C^{N}}  \otimes 
\stackrel{\stackrel{Y}{\downarrow}}{C^{2}} \,.
\ee

With respect to the all ``down'' pseudovacuum state
\be
|0\rangle_{1 \cdots L Y} =  |0\rangle_{1 \cdots L} |0\rangle_{Y} 
\,, \qquad 
|0\rangle_{1 \cdots L} =
\left( \begin{array}{c}
0 \\
\vdots\\
0 \\
1 \end{array} \right)^{\otimes L} \,,
\label{pseudovac}
\ee
$B_{j}^{(1)}(u)$ and $C_{j}^{(1)}(u)$ are annihilation and creation
operators, respectively,
\be
B_{j}^{(1)}(u) |0\rangle_{1 \cdots L Y} = 0\,, \qquad 
C_{j}^{(1)}(u) |0\rangle_{1 \cdots L Y} \ne 0\,,
\ee
and $D^{(1)}(u)$ is diagonal,
\be
D^{(1)}(u) |0\rangle_{1 \cdots L Y} =  
a(u)^{2L}\,a_{3}(u) |0\rangle_{1 \cdots L Y} \,.
\label{Dvac}
\ee
Moreover, defining the operators $\tilde A_{jk}^{(1)}(u)$ by
\be
\tilde A_{jk}^{(1)}(u) = A_{jk}^{(1)}(u) - \frac{i}{a(2u)}\delta_{jk}
D^{(1)}(u) \,,
\ee
we find that 
\be
\tilde A_{jk}^{(1)}(u) |0\rangle_{1 \cdots L Y} &=&  b(u)^{2L} 
\left[K^{-\, (1)}(u)\right]_{jk} |0\rangle_{1 \cdots L Y} \,,  
\label{tildeAvac}
\ee
where $\left[K^{-\, (1)}(u)\right]_{jk}$ are operators on the two-dimensional 
quantum space $Y$ defined by 
\be
\left[K^{-\, (1)}(u)\right]_{jk} = \left[K^{-}(u)\right]_{jk} - 
\frac{i a_{3}(u)}{a(2u)}\delta_{jk} \id \,, \qquad j, k = 1, \ldots, 
N-1 \,.
\label{KR1}
\ee

The trace over the auxiliary space in the expression 
(\ref{transfer2}) for the transfer matrix can now be performed, 
resulting in the more explicit expression
\be
t(u) = \sum_{j,k=1}^{N-1} \left[K^{+\, (1)}(u)\right]_{jk} \tilde A_{kj}^{(1)}(u)  
+ F^{(1)}(u)\, D^{(1)}(u)  
\label{transfer3} \,,
\ee
where $\left[K^{+\, (1)}(u)\right]_{jk}$ are operators on the two-dimensional 
quantum space $X$ defined by 
\be
\left[K^{+\, (1)}(u)\right]_{jk} = \left[K^{+}(u)\right]_{jk} = \left\{
\begin{array}{ll}
\alpha_{jk}(u) \,,       & j, k = 1, 2 \\
\delta_{jk} \beta(u) \,, & j, k = 3, \ldots, N-1 
\end{array} \right.
\ee
and
\be
F^{(1)}(u) = \beta(u) + \frac{i}{a(2u)}\left[ \sum_{j=1}^{2} \alpha_{jj}(u)
+ (N-3)\beta(u) \right] \,.
\label{F}
\ee 
Note that the expression (\ref{transfer3}) for $t(u)$ does {\em not}
involve either annihilation or creation operators, which is necessary
for carrying out the nested ABA analysis.

The operators obey the following commutation relations,
\be
\tilde A_{ik}^{(1)}(u)\, C_{j}^{(1)}(v) &=& 
\frac{1}{b(u-v) a(u+v)} \left[R^{(1)}(u+v+i)\right]_{i j';i' h}\, 
\left[R^{(1)}(u-v)\right]_{k' h;k j}\, C_{j'}^{(1)}(v)\, 
\tilde A_{i'k'}^{(1)}(u) \non \\
&-&\frac{i}{a(2u) b(u-v)} \left[R^{(1)}(2u+i)\right]_{i j';i' k}\, 
C_{j'}^{(1)}(u)\, 
\tilde A_{i' j}^{(1)}(v) \non \\
&+&\frac{i b(2v)}{a(2u) a(2v) a(u+v)} \left[R^{(1)}(2u+i)\right]_{i 
j';j k}\, C_{j'}^{(1)}(u)\, 
D^{(1)}(v) \,, \non \\
D^{(1)}(u)\, C_{j}^{(1)}(v) &=& 
\frac{a(v-u) b(v+u)}{b(v-u) a(v+u)} C_{j}^{(1)}(v)\, D^{(1)}(u) \non\\
&+&\frac{i b(2v)}{a(2v) b(u-v)} C_{j}^{(1)}(u)\, D^{(1)}(v)
-\frac{i}{a(u+v)} C_{j'}^{(1)}(u)\, 
\tilde A_{j' j}^{(1)}(v) \,, \non \\
C_{j}^{(1)}(u)\,  C_{k}^{(1)}(v) &=& 
\frac{1}{a(u-v)} \left[R^{(1)}(u-v)\right]_{j k; j' k'}\, C_{k'}^{(1)}(v)\,  
C_{j'}^{(1)}(u) \,,
\label{commrltns}
\ee 
where $R^{(1)}(u)$ is the $GL(N-1)$ $R$-matrix, with matrix elements
\be
\left[R^{(1)}(u)\right]_{jj;jj}&=&a(u)\,, \quad 
\left[R^{(1)}(u)\right]_{jk;jk}=b(u)\,, \quad 
\left[R^{(1)}(u)\right]_{jk;kj}=i\,, \qquad
k \ne j\,, \non \\
& & \quad j,k = 1, \ldots, N-1 \,,
\ee
with $a(u)$ and $b(u)$ as before (\ref{ab}).  Summation over repeated
indices is understood in the commutation relations.

\subsection{First level}\label{subsec:firstlevel}

The pseudovacuum state for the full space of states is given by
\be
|0\rangle_{X 1 \cdots L Y} = |0\rangle_{X} |0\rangle_{1 \cdots L Y} 
= |0\rangle_{X} |0\rangle_{1 \cdots L}  |0\rangle_{Y}
\,.
\label{omega}
\ee
It is an eigenstate of the transfer matrix (\ref{transfer3})
by virtue of (\ref{diag0vac}), (\ref{a120vac}),
(\ref{kRdiagvac}), (\ref{kR12vac}), (\ref{Dvac}) - (\ref{KR1}).
This state is {\em not} the lowest-energy state.  Indeed, it is an
eigenstate of the Hamiltonian (\ref{Hamiltonian}) with energy
eigenvalue 2, while there are eigenstates (such as the all ``up''
state) with energy 0.

We make the ansatz that the eigenstates $|\Omega^{(1)} \rangle$ of the
transfer matrix (which are independent of the spectral parameter $u$
by virtue of the commutativity property (\ref{commutativity})) can be
obtained by acting on the pseudovacuum state with the creation
operators, namely,
\be
|\Omega^{(1)} \rangle =  C_{i_{1}}^{(1)}(u_{1,1}) \ldots 
C_{i_{m_{1}}}^{(1)}(u_{1,m_{1}}) |0\rangle_{X 1 \cdots L Y} 
{\cal F}^{(1)\, i_{1}\ldots i_{m_{1}}} \,,
\label{ansatz}
\ee 
where again summation over repeated indices is understood.

By acting with the expression (\ref{transfer3}) for the transfer matrix
on this state, and using the commutation relations (\ref{commrltns}) 
to repeatedly move $\tilde A^{(1)}(u)$ and $D^{(1)}(u)$ past 
consecutive creation operators until arriving at the pseudovacuum 
state, two types of terms are generated. The  ``wanted'' terms 
are those generated by the first terms in the commutation relations;
the remaining terms are ``unwanted''. The ``wanted'' terms give
\be
t(u) |\Omega^{(1)} \rangle = \Lambda(u) |\Omega^{(1)} \rangle \,,
\ee 
with
\be
\Lambda(u) &=& f_{0}(u)\, a(u)^{2L} \prod_{j=1}^{m_{1}}
\frac{a(u_{1,j}-u) b(u_{1,j}+u)}{b(u_{1,j}-u) a(u_{1,j}+u)} \non \\
&+&  b(u)^{2L} \prod_{j=1}^{m_{1}}\frac{1}{b(u-u_{1,j}) a(u+u_{1,j})}
\Lambda^{(1)}(u\,; \left\{ u_{1,j} \right\}) \,,
\ee
where
\be
f_{0}(u) &=& a_{3}(u) \left\{b_{3}(u) + \frac{i}{a(2u)}\left[2b_{1}(u) 
+b_{2}(u) + (N-3)b_{3}(u) \right] \right\} \non \\
&=& -\frac{(2u+i N)(u^{2}+1)^{2}}{2u+i} \,.
\ee
Moreover, $\Lambda^{(1)}(u\,; \left\{ u_{1,j} \right\}) $ is a 
solution of the eigenvalue problem
\be
t^{(1)}(u\,; \left\{ u_{1,j} \right\})_{j_{1}\ldots j_{m_{1}}; i_{1}\ldots i_{m_{1}}} 
{\cal F}^{(1)\, i_{1}\ldots i_{m_{1}}} 
= \Lambda^{(1)}(u; \left\{ u_{1,j} \right\}) 
{\cal F}^{(1)\, j_{1}\ldots j_{m_{1}}} \,,
\ee
where the level-one inhomogeneous transfer matrix $t^{(1)}(u\,;
\left\{ u_{1,j} \right\})$ is defined by
\be
t^{(1)}(u\,; \left\{ u_{1,j} \right\}) = 
\tr_{a^{(1)}} K^{+\, (1)}_{a^{(1)} X}(u)\, 
{\cal T}^{-\, (1)}_{a^{(1)} 1 \cdots  m_{1} Y}(u\,;  \left\{ u_{1,j} \right\})
\label{transferlevel1}
\ee
where now the auxiliary space, denoted by $a^{(1)}$,  has dimension 
$N-1$; and
\be
{\cal T}^{-\, (1)}_{a^{(1)} 1 \cdots  m_{1} Y}(u\,;  \left\{ u_{1,j} \right\})
= T^{(1)}_{a^{(1)} 1 \cdots m_{1}}(u\,;  \left\{ u_{1,j} \right\}) 
\, K^{-\, (1)}_{a^{(1)} Y}(u)\,
\hat T^{(1)}_{a^{(1)} 1 \cdots m_{1}}(u\,;  \left\{ u_{1,j} \right\}) \,,
\ee
where the level-one inhomogeneous monodromy matrices are given by
\be
T^{(1)}_{a^{(1)} 1 \cdots m_{1}}(u\,;  \left\{ u_{1,j} \right\})  
&=& R^{(1)}_{a^{(1)} 1}(u+u_{1,1}+i) \cdots R^{(1)}_{a^{(1)} 
m_{1}}(u+u_{1,m_{1}}+i) \,, \non \\
\hat T^{(1)}_{a^{(1)} 1 \cdots m_{1}}(u\,;  \left\{ u_{1,j} \right\})
&=&  R^{(1)}_{a^{(1)} m_{1}}(u-u_{1,m_{1}}) \cdots
R^{(1)}_{a^{(1)} 1}(u-u_{1,1}) \,.
\label{monodromylevel1}
\ee 
By virtue of the fact that the level-one $K$-matrices satisfy shifted BYBEs
\be
\lefteqn{R^{(1)}_{12}(u_{1}- u_{2})\, K^{-\, (1)}_{13}(u_{1})\, 
R^{(1)}_{12}(u_{1}+u_{2}+i)\, K^{-\, (1)}_{23}(u_{2}) }\non \\
& & = K^{-\, (1)}_{23}(u_{2})\, R^{(1)}_{12}(u_{1}+u_{2}+i)\, 
K^{-\, (1)}_{13}(u_{1})\, R^{(1)}_{12}(u_{1}- u_{2}) \,,
\label{RBYBE1}
\ee 
\be
\lefteqn{R^{(1)}_{12}(-u_{1}+ u_{2})\, K^{+\, (1)}_{13}(u_{1})^{t_{1}}\, 
R^{(1)}_{12}(-u_{1}-u_{2}-\eta-i)\, 
K^{+\, (1)}_{23}(u_{2})^{t_{2}} } \label{LBYBE1} \\
& & = K^{+\, (1)}_{23}(u_{2})^{t_{2}}\, 
R^{(1)}_{12}(-u_{1}-u_{2}-\eta-i)\, 
K^{+\, (1)}_{13}(u_{1})^{t_{1}}\,R^{(1)}_{12}(-u_{1} +u_{2}) \,, 
\quad \eta = i(N-1) \,, \non
\ee 
(cf. Eqs. (\ref{RBYBE}), (\ref{LBYBE}), respectively), the level-one 
transfer matrix (\ref{transferlevel1}) has the commutativity property
\be
\left[ t^{(1)}(u\,; \left\{ u_{1,j} \right\}) \,, 
t^{(1)}(v\,; \left\{ u_{1,j} \right\}) \right] = 0 \,.
\ee
Although for the level-one transfer matrix the auxiliary space and 
the ``bulk'' quantum spaces (i.e., those labeled $1, \ldots, m_{1}$) 
have dimension one lower compared with the original transfer matrix, 
the ``boundary'' quantum spaces (i.e., those labeled $X, Y$) remain 
unchanged. 

\subsection{Iterating}\label{subsec:nesting}

We continue to iterate the above procedure. We define
\be
{\cal T}^{-\, (l)}_{a^{(l)} 1 \cdots  m_{l} Y}(u\,;  \left\{ u_{l,j} \right\})
= T^{(l)}_{a^{(l)} 1 \cdots m_{l}}(u\,;  \left\{ u_{l,j} \right\}) 
\, K^{-\, (l)}_{a^{(l)} Y}(u)\,
\hat T^{(l)}_{a^{(l)} 1 \cdots m_{l}}(u\,;  \left\{ u_{l,j} \right\}) \,,
\ee
where the auxiliary space, denoted by $a^{(l)}$, has dimension $N-l$, 
and
\be
T^{(l)}_{a^{(l)} 1 \cdots m_{l}}(u\,;  \left\{ u_{l,j} \right\})  
&=& R^{(l)}_{a^{(l)} 1}(u+u_{l,1}+i l) \cdots R^{(l)}_{a^{(l)} 
m_{l}}(u+u_{l,m_{l}}+i l) \,, \non \\
\hat T^{(l)}_{a^{(l)} 1 \cdots m_{l}}(u\,;  \left\{ u_{l,j} \right\})
&=&  R^{(l)}_{a^{(l)} m_{l}}(u-u_{l,m_{l}}) \cdots
R^{(l)}_{a^{(l)} 1}(u-u_{l,1}) \,,
\label{TTnesting}
\ee 
where $R^{(l)}(u)$ is the $GL(N-l)$ $R$-matrix,
\be
\left[R^{(l)}(u)\right]_{jj;jj}&=&a(u)\,, \quad 
\left[R^{(l)}(u)\right]_{jk;jk}=b(u)\,, \quad 
\left[R^{(l)}(u)\right]_{jk;kj}=i\,, \qquad
k \ne j\,, \non \\
& & \quad j,k = 1, \ldots, N-l \,.
\ee

We set
\be
{\cal T}^{-\, (l)}_{a^{(l)} 1 \cdots  m_{l} Y}(u\,;  \left\{ u_{l,j} \right\})  
&=& \left(
 \begin{array}{cccc}
     A_{11}^{(l+1)}(u) &\cdots & A_{1, N-l-1}^{(l+1)}(u) & B_{1}^{(l+1)}(u)\\
     \vdots    &\ddots &\vdots       & \vdots  \\
     A_{N-l-1, 1}^{(l+1)}(u) &\cdots & A_{N-l-1, N-l-1}^{(l+1)}(u) & B_{N-l-1}^{(l+1)}(u)\\
     C_{1}^{(l+1)}(u)  &\cdots & C_{N-l-1}^{(l+1)}(u)  &D^{(l+1)}(u)
 \end{array} \right)  \,, \non \\
& &\qquad l = 1, \ldots, N - 3\,, \quad N \ge 4 \,.  \label{calTform1}
\ee
The above equations are valid also for $l=0$ if we identify
\be
u_{0,j} = 0 \,, \quad m_{0} = L \,,
\ee
and also ${\cal T}^{-\, (0)} = {\cal T}^{-}$, etc., see 
(\ref{calT}),  (\ref{calTform}).
We define
\be
\tilde A_{jk}^{(l+1)}(u) = A_{jk}^{(l+1)}(u) - \frac{i}{a(2u+il)}\delta_{jk}
D^{(l+1)}(u) \,,
\ee
and find that
\be
D^{(l+1)}(u) |0\rangle_{1 \cdots m_{l} Y} &=&  
\frac{b(2u) a_{3}(u)}{b(2u+il)} \prod_{j=1}^{m_{l}} a(u-u_{l,j}) a(u+u_{l,j} + il)  
|0\rangle_{1 \cdots m_{l} Y} \,, \non \\
\tilde A_{jk}^{(l+1)}(u) |0\rangle_{1 \cdots m_{l} Y} &=& 
\prod_{j=1}^{m_{l}} b(u-u_{l,j}) b(u+u_{l,j} + il)  
\left[K^{-\, (l+1)}(u)\right]_{jk} |0\rangle_{1 \cdots m_{l} Y} \,,
\ee
where
\be
\left[K^{-\, (l+1)}(u)\right]_{jk} = \left[K^{-\, (l)}(u)\right]_{jk} - 
\frac{i b(2u) a_{3}(u)}{b(2u+i l) a(2u+i l)}\delta_{jk} \id \,, \quad j, k = 1, \ldots, 
N-l-1 \,.
\ee
The commutation relations are generalizations of (\ref{commrltns}); in
particular, the terms which generate the ``wanted'' terms are given by
\be
\tilde A_{ik}^{(l+1)}(u)\, C_{j}^{(l+1)}(v) &=& 
\frac{1}{b(u-v) a(u+v+i l)} \left[R^{(l+1)}(u+v+i(l+1))\right]_{i j';i' h} \non \\
& & \qquad \qquad \times \left[R^{(l+1)}(u-v)\right]_{k' h;k j}\, 
C_{j'}^{(l+1)}(v)\, \tilde A_{i'k'}^{(l+1)}(u) + \ldots \non \\
D^{(l+1)}(u)\, C_{j}^{(l+1)}(v) &=& 
\frac{a(v-u) b(v+u+i l)}{b(v-u) a(v+u+i l)} C_{j}^{(l+1)}(v)\, 
D^{(l+1)}(u) +\ldots  \,.
\ee

The level-$l$ transfer matrix is given by
\be
t^{(l)}(u\,; \left\{ u_{l,j} \right\}) &=& 
\tr_{a^{(l)}} K^{+\, (l)}_{a^{(l)} X}(u)\, 
{\cal T}^{-\, (l)}_{a^{(l)} 1 \cdots  m_{l} Y}(u\,;  \left\{ u_{l,j} \right\})
\non \\  
&=& \sum_{j,k=1}^{N-l-1} \left[K^{+\, (l+1)}(u)\right]_{jk} \tilde A_{kj}^{(l+1)}(u)  
+ F^{(l+1)}(u)\, D^{(l+1)}(u)  
\label{transferlevell} \,,
\ee
where 
\be
\left[K^{+\, (l+1)}(u)\right]_{jk} = \left[K^{+}(u)\right]_{jk} \,, 
\quad j, k = 1, \ldots, N-l-1 \,,
\ee
and
\be
F^{(l+1)}(u) = \beta(u) + \frac{i}{a(2u+i l)}\left[ \sum_{j=1}^{2} \alpha_{jj}(u)
+ (N-l-3)\beta(u) \right] \,.
\ee 
The $K$-matrices satisfy the shifted BYBEs
\be
\lefteqn{R^{(l)}_{12}(u_{1}- u_{2})\, K^{-\, (l)}_{13}(u_{1})\, 
R^{(l)}_{12}(u_{1}+u_{2}+i l)\, K^{-\, (l)}_{23}(u_{2}) }\non \\
& & = K^{-\, (l)}_{23}(u_{2})\, R^{(l)}_{12}(u_{1}+u_{2}+i l)\, 
K^{-\, (l)}_{13}(u_{1})\, R^{(l)}_{12}(u_{1}- u_{2}) \,,
\label{RBYBEl}
\ee 
\be
\lefteqn{R^{(l)}_{12}(-u_{1}+ u_{2})\, K^{+\, (l)}_{13}(u_{1})^{t_{1}}\, 
R^{(l)}_{12}(-u_{1}-u_{2}-\eta-i l)\, 
K^{+\, (l)}_{23}(u_{2})^{t_{2}} } \label{LBYBEl}\\
& & = K^{+\, (l)}_{23}(u_{2})^{t_{2}}\, 
R^{(l)}_{12}(-u_{1}-u_{2}-\eta-i l)\, 
K^{+\, (l)}_{13}(u_{1})^{t_{1}}\,R^{(l)}_{12}(-u_{1} +u_{2}) \,,
\quad \eta = i(N-l) \,, \non
\ee 
and therefore the level-$l$ transfer matrix also has the commutativity property.

Acting with the transfer matrix (\ref{transferlevell}) on the Bethe state
\be
|\Omega^{(l+1)} \rangle =  C_{i_{1}}^{(l+1)}(u_{l+1,1}) \ldots 
C_{i_{m_{l+1}}}^{(l+1)}(u_{l+1,m_{l+1}}) |0\rangle_{X 1 \cdots m_{l+1} Y} 
{\cal F}^{(l+1)\, i_{1}\ldots i_{m_{l+1}}} \,,
\ee 
the ``wanted'' terms give
\be
t^{(l)}(u\,; \left\{ u_{l,j} \right\}) |\Omega^{(l+1)} \rangle = 
\Lambda^{(l)}(u\,; \left\{ u_{l,j} \right\}) |\Omega^{(l+1)} \rangle \,,
\ee 
with
\be
\lefteqn{\hspace{-0.5in}\Lambda^{(l)}(u\,; \left\{ u_{l,j} \right\}) = f_{l}(u) 
\prod_{j=1}^{m_{l}}a(u-u_{l,j}) a(u+u_{l,j}+i l) 
\prod_{j=1}^{m_{l+1}}\frac{a(u_{l+1,j}-u) b(u_{l+1,j}+u+i 
l)}{b(u_{l+1,j}-u) a(u_{l+1,j}+u+i l)}} \non \\
&& \hspace{-0.5in} +\prod_{j=1}^{m_{l}}b(u-u_{l,j}) b(u+u_{l,j}+i l) 
\prod_{j=1}^{m_{l+1}}\frac{1}{b(u-u_{l+1,j}) a(u+u_{l+1,j}+i l)}
\Lambda^{(l+1)}(u\,; \left\{ u_{l+1,j} \right\}) \,,
\label{recursion}
\ee
where
\be
f_{l}(u) = f_{l}^{+}(u) f_{l}^{-}(u) \,,
\label{fl1}
\ee
and
\be
f_{l}^{-}(u)  &=& \frac{b(2u) a_{3}(u)}{b(2u+ i l)} = \frac{2u 
(u^{2}+1)}{2u+ i l} \,, \non \\
f_{l}^{+}(u)  &=& b_{3}(u) + \frac{i}{a(2u+i l)}\left[ 
2b_{1}(u)+b_{2}(u) + (N-l-3)b_{3}(u) \right] \non \\
&=& -\frac{(2u + i N)(u^{2}+1)}{2u + i (l+1)} \,, 
\qquad l = 0, \ldots, N - 3\,.
\label{fl2}
\ee

\subsection{Final level}\label{subsec:final}

We iterate the recursion relation (\ref{recursion}) until we reach 
$l=N-3$. At that stage we need the eigenvalues of the transfer matrix
$t^{(N-2)}(u\,; \left\{ u_{N-2,j} \right\}) 
= \tr_{a^{(N-2)}} K^{+\, (N-2)}(u)\, 
{\cal T}^{-\, (N-2)}(u\,;  \left\{ u_{N-2,j} \right\})$, where
the auxiliary space $a^{(N-2)}$ has only two dimensions.  The
$K$-matrices are given by
\be
K^{-\, (N-2)}(u) &=& 
\left(  \begin{array}{cccc}
a_{1}(u)+a_{2}(u) \\
& a_{1}(u) & a_{2}(u) \\
& a_{2}(u) & a_{1}(u) \\
&   &  & a_{1}(u)+a_{2}(u) \\
\end{array} \right) - \frac{i (N-2) a_{3}(u)}{2u+i(N-2)}\id \,,  \non \\
K^{+\, (N-2)}(u) &=&  
\left(  \begin{array}{cccc}
b_{1}(u)+b_{2}(u) \\
& b_{1}(u) & b_{2}(u) \\
& b_{2}(u) & b_{1}(u) \\
&   &  & b_{1}(u)+b_{2}(u) \\
\end{array} \right) \,,
\label{KK2}
\ee
where matrix elements which are zero are left empty. They obey the 
shifted BYBEs (\ref{RBYBEl}), (\ref{LBYBEl}), respectively, with $l=N-2$.

A priori, one would expect to encounter serious difficulty in
diagonalizing this transfer matrix, since both $K$-matrices (in
particular, the left one) are not diagonal.  Remarkably, this is not
the case.  Indeed, we note the identity
\be
\lefteqn{t^{(N-2)}(u\,; \left\{ u_{N-2,j} \right\}) }\non \\ 
& & = -\frac{2u}{2u+i(N-2)}\tr_{a} 
S^{(N-2)}_{a X 1 \cdots m_{N-2} Y}(u\,;  \left\{ u_{N-2,j} \right\})\,  
\hat S^{(N-2)}_{a X 1 \cdots m_{N-2} Y}(u\,;  \left\{ u_{N-2,j} 
\right\}) \,, 
\label{remarkable2}
\ee
where
\be
S^{(N-2)}_{a X 1 \cdots m_{N-2} Y}(u\,;  \left\{ u_{N-2,j} \right\})
&=& R^{(N-2)}_{a X}(u +i(N-2))\, 
R^{(N-2)}_{a 1}(u+u_{N-2,1}+i(N-2)) \cdots  \non  \\
&\times & 
R^{(N-2)}_{a m_{N-2}}(u+u_{N-2,m_{N-2}}+i(N-2))\,
R^{(N-2)}_{a Y}(u +i(N-2))\,, \non \\
\hat S^{(N-2)}_{a X 1 \cdots m_{N-2} Y}(u\,;  \left\{ u_{N-2,j} \right\})
&=& R^{(N-2)}_{a Y}(u)\,
R^{(N-2)}_{a m_{N-2}}(u-u_{N-2,m_{N-2}})\, \cdots \non\\
& & \times \quad 
R^{(N-2)}_{a 1}(u-u_{N-2,1})\,
R^{(N-2)}_{a X}(u) \,, \label{TT2}
\ee
and $a \equiv a^{(N-2)}$ is the two-dimensional auxiliary space.  That
is, the transfer matrix $t^{(N-2)}(u\,; \left\{ u_{N-2,j} \right\})$
is the same as the transfer matrix of an open inhomogeneous spin-1/2
$GL(2)$-invariant chain of length $2+m_{N-2}$ with trivial
$K$-matrices (i.e., equal to the identity matrix).\footnote{A similar
observation (although without proof and only for the case $M=2$) has
been made for related models in \cite{ZG, ZGLG, ZGLG2}.} A proof for
general values of $M$ is given in Appendix \ref{sec:proof}.  The
corresponding eigenvalues can therefore be determined by standard
methods such as \cite{Sk}, and we obtain
\be
\hspace{-0.3in}\Lambda^{(N-2)}(u\,; \left\{ u_{N-2,j} \right\})  &=&
f_{N-2}(u) \prod_{j=1}^{m_{N-2}}(u-u_{N-2,j}+i)(u+u_{N-2,j}+i(N-1)) 
\non \\
& & \times \prod_{j=1}^{m_{N-1}}
\frac{(u-u_{N-1,j}-i)(u+u_{N-1,j}+i(N-2))}
{(u-u_{N-1,j})(u+u_{N-1,j}+i(N-1))}
\non \\
&+&
f_{N-1}(u) \prod_{j=1}^{m_{N-2}}(u-u_{N-2,j})(u+u_{N-2,j}+i(N-2)) 
\non \\
& & \times
\prod_{j=1}^{m_{N-1}} 
\frac{(u-u_{N-1,j}+i)(u+u_{N-1,j}+i N)}
{(u-u_{N-1,j})(u+u_{N-1,j}+i(N-1))}
\,,
\ee
where
\be
f_{N-2}(u) &=& -\frac{2u(u+i)^{2}(u+i(N-1))^{2}(2u+i N)}{(2u+i(N-2))(2u+i(N-1))}\,, \non \\
f_{N-1}(u) &=& -\frac{2u^{3}(u+i(N-2))^{2}}{2u+i(N-1)}\,.
\label{fl3}
\ee 

Combining the above results, we conclude that the eigenvalues of the 
original transfer matrix (\ref{transfer}) with $M=2$ and $N \ge 3$ are given by
\be
\Lambda(u) &=& f_{0}(u) (u+i)^{2L} 
\frac{Q_{1}(u-\frac{i}{2})}{Q_{1}(u+\frac{i}{2})}
+ u^{2L} \Bigg\{ \sum_{l=1}^{N-2} f_{l}(u)
\frac{Q_{l}(u+\frac{i}{2}(l+2))}{Q_{l}(u+\frac{i}{2}l)}
\frac{Q_{l+1}(u+\frac{i}{2}(l-1))}{Q_{l+1}(u+\frac{i}{2}(l+1))}\non \\
&+&  f_{N-1}(u)
\frac{Q_{N-1}(u+\frac{i}{2}(N+1))}{Q_{N-1}(u+\frac{i}{2}(N-1))} 
\Bigg\} \,,
\label{eigenvals}
\ee
where
\be
Q_{l}(u) = \prod_{j=1}^{m_{l}} (u - u_{l, j}) (u + u_{l, j}) \,, 
\qquad l = 1, \ldots , N-1 \,,
\label{Qs}
\ee
and we have made the shifts $u_{l,j} \mapsto u_{l,j} - \frac{i}{2}l$. 
We recall that the functions $f_{l}(u)$ are given by (\ref{fl1}),
(\ref{fl2}), (\ref{fl3}).

We have thus far ignored all the contributions from ``unwanted'' terms in 
the commutation relations. Such contributions vanish provided the 
parameters $\{ u_{l,j} \}$ satisfy the Bethe ansatz equations
\be
e_{1}(u_{1, k})^{2L}  &=&  \Theta_{1}(u_{1, k})
\prod_{j=1 \atop j\ne k}^{m_{1}} e_{2}(u_{1, k} - u_{1, j})\, 
e_{2}(u_{1, k} + u_{1, j}) \non \\
& & \times \prod_{j=1}^{m_{2}} e_{-1}(u_{1, k} - 
u_{2, j})\, e_{-1}(u_{1, k} + u_{2, j}) \,, \qquad 
k = 1\,, \ldots \,, m_{1} \,, \non \\
1 &=& \Theta_{l}(u_{l, k})
\prod_{j=1 \atop j\ne k}^{m_{l}} e_{2}(u_{l, k} - u_{l, j})\, 
e_{2}(u_{l, k} + u_{l, j})
\prod_{j=1}^{m_{l-1}} e_{-1}(u_{l, k} - 
u_{l-1, j})\, e_{-1}(u_{l, k} + u_{l-1, j}) \non \\
&\times &  \prod_{j=1}^{m_{l+1}} e_{-1}(u_{l, k} - 
u_{l+1, j})\, e_{-1}(u_{l, k} + u_{l+1, j})\,,  \quad k = 1\,, \ldots 
\,, m_{l} \,, \quad l = 2, \ldots, N-2\,, \non \\
1 &=& \Theta_{N-1}(u_{N-1, k})
\prod_{j=1 \atop j\ne k}^{m_{N-1}} e_{2}(u_{N-1, k} - u_{N-1, j})\, 
e_{2}(u_{N-1, k} + u_{N-1, j})\non \\
& & \times  \prod_{j=1}^{m_{N-2}} e_{-1}(u_{N-1, k} - 
u_{N-2, j})\, e_{-1}(u_{N-1, k} + u_{N-2, j}) \,,  \quad k = 1\,, \ldots 
\,, m_{N-1}\,,
\label{BAEs}
\ee 
where 
\be
 \Theta_{l}(u) = \left\{ \begin{array}{cl} 
 e_{N}(u)^{2} & \qquad l = N-2 \\ \\
 \frac{e_{N-3}(u)^{2}}{e_{N-1}(u)^{2}} & \qquad l = N-1   \\ \\
 1  & \qquad  \mbox{otherwise}
 \end{array}\right. \,,
\ee
and we have used the standard notation
\be
e_{n}(u) = \frac{u + i n/2}{u - i n/2} \,.
\label{notation}
\ee

Finally, from the relation (\ref{tHrelation}) between the transfer
matrix and the Hamiltonian, we find that the energy eigenvalues are
given by
\be
E = c_{1} \frac{d}{du} \Lambda(u) \Big\vert_{u=0} + c_{2} 
  = 2 + \sum_{k=1}^{m_{1}} \frac{1}{u_{1, k}^{2} + 1/4} \,.
\label{energy}
\ee

\subsection{The case $N=3\,, M=2$}\label{subsec:N3}

For the case $(N,M)=(3,2)$, the above results do {\em not} coincide with 
those in our previous work \cite{Ne2}. Indeed, there we found that 
the eigenvalues are given by the same expression (\ref{eigenvals})
but with different functions $f_{l}(u)$, namely,
\be
f_{0}^{\textrm{previous}}(u) &=& - \frac{(2u+3i)(u+i)^{4}}{2u+i} 
= \left(\frac{u+i}{u-i}\right)^{2} f_{0}(u) \,, \non \\
f_{1}^{\textrm{previous}}(u) &=& - \frac{u^{3}(2u+3i)(u+i)}{2u+i}
 = \left(\frac{u}{u+2i}\right)^{2} f_{1}(u) \,, \non \\
f_{2}^{\textrm{previous}}(u) &=& - u^{3}(u+i) =  f_{2}(u) \,. 
\ee
(See Eqs. (2.33) and (2.36) in \cite{Ne2}.) Equivalently, the two 
sets of results can instead be related by
\be
Q_{1}(u) &=& g(u)\, Q_{1}^{\textrm{previous}}(u) \,, \quad 
g(u) =(u +\frac{i}{2})^{-2} (u -\frac{i}{2})^{-2} \,, \non \\
Q_{2}(u) &=&  Q_{2}^{\textrm{previous}}(u) \,,
\ee 
since
\be
\frac{g(u-\frac{i}{2})}{g(u+\frac{i}{2})} = 
\frac{f_{0}^{\textrm{previous}}(u)}{f_{0}(u)} \,, \qquad
\frac{g(u+\frac{3i}{2})}{g(u+\frac{i}{2})} = 
\frac{f_{1}^{\textrm{previous}}(u)}{f_{1}(u)} \,.
\ee

The discrepancy in the two sets of results arises from different
choices of pseudovacua.  In \cite{Ne2} we chose the pseudovacuum to be
a ground state $(E = 0)$, while here we have taken the pseudovacuum to
be an excited state $(E = 2)$. (Notice the additive constant in the 
expression (\ref{energy}) for the energy.)

We have performed a numerical analysis of completeness of the new
solution for small values of $L$ along the lines discussed in Appendix
B of \cite{Ne2}.  The results for the case $L=3$, for which case there
are $M^{2} N^{L} = 108$ states, are displayed in Table \ref{table:N3}.
Although we find some levels for which $m_{2} > m_{1}$ (which we did
not find with our previous solution), this solution also appears to be
complete, at least for small values of $L$.  Note that the Bethe roots
for the ground ($E=0$) state have a rather complicated structure.
Comparing this table with Table 2 in Ref.  \cite{Ne2}, we see little
apparent relation between the two sets of Bethe roots describing a
given energy level.

It would be interesting to re-derive our previous solution \cite{Ne2}
(obtained by analytic Bethe ansatz, which is a heuristic approach) by
the more rigorous nested ABA approach considered here.  Unfortunately,
we have so far not succeeded.  Indeed, if we try to use the all ``up''
state as the pseudovacuum, then the creation and annihilation
operators seem to be $A^{(1)}_{12}(u)\,, B^{(1)}_{1}(u)$ and
$A^{(1)}_{21}(u)\,, C^{(1)}_{1}(u)$, respectively; hence the transfer
matrix seems to involve creation and annihilation operators.

\subsection{The cases $N > 3\,, M=2$}\label{subsec:N4}

For $N > 3\,, M=2$, the solution also seems to be complete.  For
example, we display in Table \ref{table:N4} our results for
$(N,M)=(4,2)$ and $L=2$, for which case there are $M^{2} N^{L} = 64$
states.

\section{Nested ABA for general values of $M$}\label{sec:nestedM}

For $M\ge 2$, the left $K$-matrix has the form (cf. Eq. (\ref{KLform}))
\be
K^{+}_{a X}(u) = \left(
 \begin{array}{cccccc}
     \alpha_{11}(u) & \cdots & \alpha_{1M}(u) & 0 &\cdots & 0\\
     \vdots         & \ddots & \vdots         & \vdots &\ddots & 
     \vdots \\
     \alpha_{M1}(u) & \cdots & \alpha_{MM}(u) & 0 &\cdots & 0\\
	0           & \cdots &   0            & \beta(u) &\cdots & 0\\
     \vdots    & \ddots  & \vdots  &\vdots &\ddots  & \vdots  \\
     	0 & \cdots & 0  & 0 &\cdots & \beta(u)
 \end{array} \right)  \,.
\ee
We can therefore iterate the recursion relation (\ref{recursion})
until we reach $l=N-M-1$.  At that stage we need the eigenvalues of
the transfer matrix $t^{(N-M)}(u\,; \left\{ u_{N-M,j} \right\})$, for
which the auxiliary space $a^{(N-M)}$ has dimension $M$.  The
corresponding  $K$-matrices are given by
\be
K^{-\, (N-M)}(u) &=& \left[a_{1}(u) -\frac{i (N-M) 
a_{3}(u)}{2u+i(N-M)} \right] \id
 + a_{2}(u)\, {\cal P} \,, \non \\
K^{+\, (N-M)}(u) &=& b_{1}(u)\, \id 
+ b_{2}(u)\, {\cal P} \,,
\label{KM}
\ee
where $\id$ and ${\cal P}$ are the identity and permutation matrices  
on $C^{M} \otimes C^{M}$, respectively.
They obey the shifted BYBEs (\ref{RBYBEl}), (\ref{LBYBEl}),
respectively, with $l=N-M$.

Since these $K$-matrices (in particular, the left one) are not
diagonal, it is not evident how to diagonalize the transfer matrix.
Fortunately, there is an identity generalizing (\ref{remarkable2}),
(\ref{TT2}), namely
\be
\lefteqn{t^{(N-M)}(u\,; \left\{ u_{N-M,j} \right\})}\non \\ 
& & = -\frac{2u}{2u+i(N-M)}\tr_{a} 
S^{(N-M)}_{a X 1 \cdots m_{N-M} Y}(u\,;  \left\{ u_{N-M,j} \right\})\,  
\hat S^{(N-M)}_{a X 1 \cdots m_{N-M} Y}(u\,;  \left\{ u_{N-M,j} 
\right\}) \,, 
\label{remarkableM}
\ee
where
\be
S^{(N-M)}_{a X 1 \cdots m_{N-M} Y}(u\,;  \left\{ u_{N-M,j} \right\})
&=& R^{(N-M)}_{a X}(u +i(N-M))\, 
R^{(N-M)}_{a 1}(u+u_{N-M,1}+i(N-M)) \cdots \non \\
&\times & 
R^{(N-M)}_{a m_{N-M}}(u+u_{N-M,m_{N-M}}+i(N-M))\,
R^{(N-M)}_{a Y}(u +i(N-M))\,, \non \\
\hat S^{(N-M)}_{a X 1 \cdots m_{N-M} Y}(u\,;  \left\{ u_{N-M,j} \right\})
&=& R^{(N-M)}_{a Y}(u)\,
R^{(N-M)}_{a m_{N-M}}(u-u_{N-M,m_{N-M}})\, \cdots \non\\
& & \times \quad 
R^{(N-M)}_{a 1}(u-u_{N-M,1})\,
R^{(N-M)}_{a X}(u) \,, \label{TTM}
\ee
and $a \equiv a^{(N-M)}$ is the $M$-dimensional auxiliary space.  That
is, the transfer matrix is the same as that of an open inhomogeneous
$GL(M)$-invariant chain of length $2+m_{N-M}$ with trivial
$K$-matrices and spins in the vector ($M$-dimensional) representation.
See Appendix \ref{sec:proof} for a proof.  The corresponding
eigenvalues can be found by the ``ordinary'' nested ABA \cite{FK,
dVGR}, and we obtain
\be
\lefteqn{\Lambda^{(N-M)}(u\,; \left\{ u_{N-M,j} \right\}) = 
f_{N-M}(u) \prod_{j=1}^{m_{N-M}}(u-u_{N-M,j}+i)(u+u_{N-M,j}+i(N-M+1)) }
\non \\
& & \qquad \qquad \qquad \qquad \qquad \times \prod_{j=1}^{m_{N-M+1}}
\frac{(u-u_{N-M+1,j}-i)(u+u_{N-M+1,j}+i(N-M))}
{(u-u_{N-M+1,j})(u+u_{N-M+1,j}+i(N-M+1))}
\non \\
& & +
\prod_{j=1}^{m_{N-M}} (u-u_{N-M,j})(u+u_{N-M,j}+i(N-M)) \non \\
& & \times \Bigg[\sum_{l=N-M+1}^{N-2}f_{l}(u)
\prod_{j=1}^{m_{l}} 
\frac{(u-u_{l,j}+i)(u+u_{l,j}+i(l+1))}{(u-u_{l,j})(u+u_{l,j}+i l)}
\prod_{j=1}^{m_{l+1}} 
\frac{(u-u_{l+1,j}-i)(u+u_{l+1,j}+i 
l)}{(u-u_{l+1,j})(u+u_{l+1,j}+i(l+1))}\non \\
& & \qquad \qquad + f_{N-1}(u)  \prod_{j=1}^{m_{N-1}} 
\frac{(u-u_{N-1,j}+i)(u+u_{N-1,j}+i N)}
{(u-u_{N-1,j})(u+u_{N-1,j}+i(N-1))} \Bigg]
\,,
\ee
where 
\be
f_{l}(u) =  \left\{
\begin{array}{ll}
    -\frac{2u(u+i)^{2}(u+i(N-M+1))^{2}(2u+i 
    N)}{(2u+i(N-M))(2u+i(N-M+1))} & l = N-M \\ \\
    -\frac{2u^{3}(u+i(N-M))^{2}(2u+i 
    N)}{(2u+i l)(2u+i(l+1))} & l = N-M+1, \ldots, N-1 
    \end{array}\right. \,.
\ee

Combining this result with those from the recursion relation 
(\ref{recursion}), we conclude that the eigenvalues of the 
original transfer matrix (\ref{transfer}) are given by
\be
\Lambda(u) &=& f_{0}(u) (u+i)^{2L} 
\frac{Q_{1}(u-\frac{i}{2})}{Q_{1}(u+\frac{i}{2})}
+ u^{2L} \Bigg\{ \sum_{l=1}^{N-2} f_{l}(u)
\frac{Q_{l}(u+\frac{i}{2}(l+2))}{Q_{l}(u+\frac{i}{2}l)}
\frac{Q_{l+1}(u+\frac{i}{2}(l-1))}{Q_{l+1}(u+\frac{i}{2}(l+1))}\non \\
&+&  f_{N-1}(u)
\frac{Q_{N-1}(u+\frac{i}{2}(N+1))}{Q_{N-1}(u+\frac{i}{2}(N-1))} 
\Bigg\} \,,
\label{eigenvalsM}
\ee
where
\be
f_{l}(u) = \left\{
\begin{array}{ll}
    -\frac{2u(u^{2}+1)^{2}(2u+i N)}{(2u+i l)(2u+i(l+1))} & l = 0, 
    \ldots, N-M-1 \\ \\
   -\frac{2u(u+i)^{2}(u+i(N-M+1))^{2}(2u+i 
    N)}{(2u+i(N-M))(2u+i(N-M+1))} & l = N-M \\ \\
    -\frac{2u^{3}(u+i(N-M))^{2}(2u+i 
    N)}{(2u+i l)(2u+i(l+1))} & l = N-M+1, \ldots, N-1 
    \end{array}\right. \,,
\ee
\be
Q_{l}(u) = \prod_{j=1}^{m_{l}} (u - u_{l, j}) (u + u_{l, j}) \,, 
\qquad l = 1, \ldots , N-1 \,,
\label{QsM}
\ee 
and (as before) we have made the shifts $u_{l,j} \mapsto u_{l,j} - \frac{i}{2}l$.

The corresponding Bethe ansatz equations are given by
\be
e_{1}(u_{1, k})^{2L}  &=& \Theta_{1}(u_{1, k})
\prod_{j=1 \atop j\ne k}^{m_{1}} e_{2}(u_{1, k} - u_{1, j})\, 
e_{2}(u_{1, k} + u_{1, j}) \non \\
& & \times \prod_{j=1}^{m_{2}} e_{-1}(u_{1, k} - 
u_{2, j})\, e_{-1}(u_{1, k} + u_{2, j}) \,, \qquad 
k = 1\,, \ldots \,, m_{1} \,, \non \\
1 &=& \Theta_{l}(u_{l, k} )
\prod_{j=1 \atop j\ne k}^{m_{l}} e_{2}(u_{l, k} - u_{l, j})\, 
e_{2}(u_{l, k} + u_{l, j})
\prod_{j=1}^{m_{l-1}} e_{-1}(u_{l, k} - 
u_{l-1, j})\, e_{-1}(u_{l, k} + u_{l-1, j}) \non \\
&\times &  \prod_{j=1}^{m_{l+1}} e_{-1}(u_{l, k} - 
u_{l+1, j})\, e_{-1}(u_{l, k} + u_{l+1, j})\,,  \quad k = 1\,, \ldots 
\,, m_{l} \,, \quad l = 2, \ldots, N-2\,, \non \\
1 &=& \Theta_{N-1}(u_{N-1, k})
\prod_{j=1 \atop j\ne k}^{m_{N-1}} e_{2}(u_{N-1, k} - u_{N-1, j})\, 
e_{2}(u_{N-1, k} + u_{N-1, j})\non \\
& & \times  \prod_{j=1}^{m_{N-2}} e_{-1}(u_{N-1, k} - 
u_{N-2, j})\, e_{-1}(u_{N-1, k} + u_{N-2, j}) \,,  \quad k = 1\,, \ldots 
\,, m_{N-1}\,,
\label{BAEsM}
\ee 
where now
\be
 \Theta_{l}(u) = \left\{ \begin{array}{cl} 
 e_{N-M+2}(u)^{2} & \qquad l = N-M \\ \\
 \frac{e_{N-M-1}(u)^{2}}{e_{N-M+1}(u)^{2}} & \qquad l = N-M+1   \\ \\
 1  & \qquad  \mbox{otherwise}
 \end{array}\right. \,,
 \label{Theta}
\ee
and $e_{n}(u)$ is defined in (\ref{notation}).
The energy eigenvalues are given by the same formula (\ref{energy}).
The identity (\ref{remarkableM}), the expression
(\ref{eigenvalsM})-(\ref{QsM}) for the eigenvalues of the transfer
matrix and the corresponding Bethe ansatz equations (\ref{BAEsM}),
(\ref{Theta}) are the main results of this paper.

For $M>2$, this solution also seems to be complete, as is the case
for $M=2$ discussed in Sec.  \ref{subsec:N4}.  For example,
we display in Table \ref{table:M3} our results for $(N,M)=(4,3)$ and
$L=2$, for which case there are $M^{2} N^{L} = 144$ states.  

\section{Conclusions}\label{sec:conclusion}

We have considered the $GL(N)/(GL(M) \times GL(N-M))$ model
with Hamiltonian (\ref{Hamiltonian}), which is a generalization of a
model arising in string/gauge theory.  We have proved the
integrability of this model by constructing the corresponding
commuting transfer matrix.  The latter makes use of the non-diagonal
operator-valued $K$-matrices found in \cite{FS}.

We have found a Bethe ansatz solution of this model for general values
of $N$ and $M$ using the nested ABA approach, despite the fact that
the $K$-matrices are not diagonal.  The main results are the
eigenvalues (\ref{eigenvalsM})-(\ref{QsM}) and Bethe ansatz equations
(\ref{BAEsM}), (\ref{Theta}).  The key to obtaining this solution is
the identity (\ref{remarkableM}), which relies on the factorization
property (\ref{factorization}) of the ``reduced'' (level $N-M$)
$K$-matrices into products of $R$-matrices.  In hindsight, this
property is not too surprising, since the projected $K$-matrices
originate from ``dressed'' diagonal $K$-matrices \cite{FS}.  For the
case $(N,M) = (3,2)$, this solution is not the same as the one found
in \cite{Ne2} using analytic Bethe ansatz, as the two solutions are
based on different pseudovacua.  Nevertheless, numerical evidence
suggests that both $N=3$ solutions are complete.  Moreover, the nested
ABA solution appears to be complete for general values of $N$ and $M$.

Many interesting questions remain unanswered.  It is unusual for an
integrable model with a non-graded symmetry algebra to have more than
one Bethe ansatz solution.  (Models with graded symmetry algebras are
known to have more than one Bethe ansatz solution, corresponding to
the non-uniqueness of the associated Dynkin diagrams.  See e.g.
\cite{duality} and references therein.)  This underscores the question
of whether the two proposed solutions for the case $(N,M) = (3,2)$
(namely, the one found in \cite{Ne2} by analytic Bethe ansatz, and the
one found here by nested ABA) are equivalent.  As noted in Sec.
\ref{subsec:N3}, one would like to have a more rigorous derivation of
the solution found in \cite{Ne2}.  Similarly, for general values of
$N$ and $M$, there may be additional equivalent solutions based on
different pseudovacua.  Perhaps Bethe ansatz equations for generic
open spin chains (or at least for open chains constructed with
projected $K$-matrices) can be formulated in terms of group theory
data (namely, the ``bulk'' symmetry algebra and the unbroken
``boundary'' symmetry subalgebra); and the multiplicity of Bethe
ansatz solutions reflects the various ways of choosing the boundary
symmetry subalgebra.  We hope to be able to address these and related
questions in the future.

\section*{Acknowledgments}
I am grateful to A. Lima-Santos, J. Links and M. Martins for helpful 
correspondence.
This work was supported in part by the National Science
Foundation under Grants PHY-0554821 and PHY-0854366.

\appendix

\section{Proof of the transfer-matrix identity}\label{sec:proof}

Our proof of the transfer-matrix identity (\ref{remarkable2}),
(\ref{remarkableM}) is based on the following remarkable factorization
property of the ``reduced'' $K$-matrices (i.e., the $(N,M)$ projected
$K$-matrices at level $N-M$) into products of $R$-matrices,
\be
K^{-\, (N-M)}_{a Y}(u) &=& -\frac{2u}{2u+i(N-M)}
R^{(N-M)}_{a Y}(u +i(N-M))\, R^{(N-M)}_{a Y}(u) \,, \non \\
K^{+\, (N-M)}_{a X}(u) &=& \tr_{b} {\cal P}_{a b}\,
R^{(N-M)}_{a X}(u +i(N-M))\, R^{(N-M)}_{b X}(u) \,,
\label{factorization}
\ee
which can be verified from the expressions (\ref{KK2}), (\ref{KM}).
Omitting the quantum-space indices and denoting the $M$-dimensional
auxiliary space by $a$ in order to streamline the notation, we have
\be
t^{(N-M)}(u) &=& \tr_{a} K^{+\, (N-M)}_{a}(u)\, {\cal T}^{-\, 
(N-M)}_{a}(u) \non  \\
 &=&  \tr_{a b} {\cal P}_{a b}\,
R^{(N-M)}_{a}(u +i(N-M))\, R^{(N-M)}_{b}(u)\, {\cal T}^{-\, 
(N-M)}_{a}(u) \non \\
&=&  \tr_{a b} {\cal P}_{a b}\,
R^{(N-M)}_{a}(u +i(N-M))\,  {\cal T}^{-\, 
(N-M)}_{a}(u)\, R^{(N-M)}_{b}(u) \non \\
&=&  \tr_{a} 
R^{(N-M)}_{a}(u +i(N-M))\,  {\cal T}^{-\, 
(N-M)}_{a}(u)\, R^{(N-M)}_{a}(u) \non \\
&=&  \tr_{a} 
R^{(N-M)}_{a}(u +i(N-M))\,  T^{(N-M)}_{a}(u)\, 
K^{-\, (N-M)}_{a}(u)\, \hat T^{(N-M)}_{a}(u)\, 
R^{(N-M)}_{a}(u) \non\\
&=&  -\frac{2u}{2u+i(N-M)} \tr_{a} 
R^{(N-M)}_{a}(u +i(N-M))\,  T^{(N-M)}_{a}(u) \non \\ 
& & \times \quad R^{(N-M)}_{a}(u +i(N-M))\, R^{(N-M)}_{a}(u)\, 
\hat T^{(N-M)}_{a}(u)\, 
R^{(N-M)}_{a}(u) \non \\
&=&  -\frac{2u}{2u+i(N-M)} \tr_{a} 
S^{(N-M)}_{a}(u)\,  
\hat S^{(N-M)}_{a}(u) \,,
\ee
where in the last line we have used the fact (see Eqs. (\ref{TTM}), 
(\ref{TTnesting}))
\be
S^{(N-M)}_{a}(u) &=&  R^{(N-M)}_{a}(u +i(N-M))\,  
T^{(N-M)}_{a}(u)\, R^{(N-M)}_{a}(u +i(N-M)) \,, \non \\
\hat S^{(N-M)}_{a}(u)  &=& R^{(N-M)}_{a}(u)\,
\hat T^{(N-M)}_{a}(u)\, R^{(N-M)}_{a}(u) \,.
\ee

\newpage
\begin{table}[h!] 
  \centering
  \begin{tabular}{|c|c|c|c|c|}\hline
     $E$ & $s$ & $\{ u_{1, k} \}$ & $\{ u_{2, k} \}$\\
    \hline
          0  &  5/2  & $1.11803 i\,, 0.442686 \pm 1.0936 i$  & -- \\
         0.381966 & 3/2 & $1.2944 i\,, 0.375279 \pm 1.36374 i$  & 0  \\
	 0.585786 & 2 & $0.204205 \pm 1.22426 i$ & -- \\
	 0.82259  & 1/2, 1 & $0.15313 \pm 1.36461 i$ & 0 \\
	 1.07919  & 0 & $1.36676 i\,, 1.88488 i$ & $0\,, 1.56857 i$ \\
	 1.26795  & 3/2 & $1.27123 i$ & -- \\
	 1.38197  & 3/2 & $0.936268\,, 0.180565 \pm 1.20371 i$  & 0 \\
	 1.38197  & 1/2 & $1.36676 i$ & 0 \\
	 1.58579  & 1/2, 1 & $1.91214\,, 1.31987 i$ & 0  \\
	 1.69722  & 1/2 & $1.88488 i$ &  0 \\
	 2        & 2 & $0.866025\,, 1.11803 i$ & -- \\
	 2        & 1, 3/2 & -- & -- \\
	 2        & 0, 1  & -- & 0 \\
	 2.58579  & 1    & $0.639467 \,, 1.15027 i$ & 1.0505 \\
	 2.61803  & 3/2 & $0.322878 \pm 0.500421 i\,, 1.04607 i$ & 0 \\
	 3        & 0 & $0.606658\,, 1.36676 i$ & $0 \,, 0.707107 i$ \\
	 3        & 1/2 & 0.866025 &  1 \\
	 3.31526  & 0 & $0.606658\,, 1.88488 i$ & $1.15861 i$ \\
	 3.32164  & 1/2, 1 & $0.451092\,, 1.17552 i$ & 0  \\
	 3.41421  & 2 & $0.479032 \pm 0.521886 i$ & -- \\
	 3.61803  & 1/2 & 0.606658 & 0 \\
	 3.61803  & 3/2 & $0.331608\,, 0.404442 \pm 0.90768 i$  & 0 \\
	 4        & 1 & $ \pm 0.5 i$ &  1 \\
	 4.41421  & 1/2, 1 & $0.0774471\,, 0.959277 i$ & 0 \\
	 4.68474  & 0 & $0.229729\,, 1.36676 i$ & $0 \,, 0.810943 i$ \\
	 4.73205  & 3/2 & 0.340625 & -- \\
	 5        & 1/2 & 0.288675  & 0.745356 \\
	 5        & 0 & $0.229729\,, 1.88488 i$ & $0\,, 1.22474 i$ \\
	 5.30278  & 1/2 & 0.229729 & 0 \\
	 5.41421  & 1 & $0.301797\,, 1.35023 $ & 0.62964 \\
	 5.85577  & 1/2, 1 & $0.248411\,, 1.13757$ & 0 \\
	 6.92081  & 0 & $0.229729\,, 0.606658 $  & $0\,, 0.678531$ \\
         \hline
   \end{tabular}
   \caption{Energy, spin, and Bethe 
   roots for $N = 3\,, M =2\,, L = 3$.}
   \label{table:N3}
\end{table}

\strut

\newpage
\strut

\begin{table}[h!] 
  \centering
  \begin{tabular}{|c|c|c|c|c|c|}\hline
     $E$ & $deg$ & $\{ u_{1, k} \}$ & $\{ u_{2, k} \}$ & $\{ u_{3, k} 
     \}$ \\
    \hline
          0  & 5   & $0.238862 \pm 0.986773 i$  & $0.240994 \pm 1.54642 
	  i$ & -- \\
	  0.585786 & 3 & $0.204205 \pm 1.22426 i$ & $\pm 1.84776 i$ & 
	  $1.39897 i$ \\
	  1  & 8   & $1.11803 i $ & $ 1.63299 i$ & -- \\
	  1.26795 & 5 & $1.27123 i$ & $1.79779 i$ & $1.01915 i$ \\
	  2  & 9   &  -- & -- & -- \\
	  2  & 7   &  -- & -- & $0.866025 i$ \\
	  2  & 3   & $0.866025\,, 1.11803 i$ & $0.8556\,, 1.65289 i$ 
	  & $0.866025 i$ \\
	  3  & 8   & 0.866025 & 0 & -- \\
	  3.41421 & 3 & $0.479032 \pm 0.521886 i$ & $\pm 0.765367 i$ & 
	  $0.736813 i$ \\
	  4  & 7   & 0.5  & -- & -- \\
	  4  & 1   & 0.5  & -- & $0.866025 i$ \\
	  4.73205 & 5 & 0.340625 & 0.481717 & $0.679209 i$ \\	  
         \hline
   \end{tabular}
   \caption{Energy, degeneracy, and Bethe 
   roots for $N = 4\,, M =2\,, L = 2$.}
   \label{table:N4}
\end{table}

\strut


%
\begin{table}[h!] 
  \centering
  \begin{tabular}{|c|c|c|c|c|c|}\hline
     $E$ & $deg$ & $\{ u_{1, k} \}$ & $\{ u_{2, k} \}$ & $\{ u_{3, k} 
     \}$ \\
    \hline
       0       & 15 & $0.238862 \pm 0.986773 i$ & -- & -- \\
       0.585786 & 15 & $0.204205 \pm 1.22426 i$ & 0 & -- \\
       1       & 10 & $1.11803 i$ & -- & -- \\
       1.26795 & 14 & $1.27123 i$ & 0 & -- \\
       2       &  6 & --       & --   & -- \\
       2       & 11 & --       & 0    & -- \\
       2       & 15 & $0.866025\,, 1.11803 i$ & 0 & -- \\
       2.58579 &  3 & $0.639467\,, 1.15027 i$  & $0\,, 1.0505  $ & 0.89542  \\
       3       & 10 & 0.866025 & -- & -- \\
       3       &  1 & 0.866025 & 0, 1     & 0.866025 \\
       3.41421 & 15 & $0.479032 \pm 0.521886 i$ & 0 & -- \\
       4       &  8 & 0.5      & 0.816497 & -- \\
       4       &  3 & $ \pm 0.5 i$        & $0\,, 1$  & 0.866025  \\
       4.73205 & 14 & 0.340625 & 0 & -- \\
       5       &  1 & 0.288675 & 0, 0.745356 & 0.726483 \\ 
       5.41421 &  3 & $0.301797\,, 1.35023 $       & $0\,, 0.62964 $  & 0.669495  \\
    \hline
   \end{tabular}
   \caption{Energy, degeneracy, and Bethe 
   roots for $N = 4\,, M =3\,, L = 2$.}
   \label{table:M3}
\end{table}

\strut


\end{document}